\documentclass[twocolumn,journal]{IEEEtran}

\usepackage{amsmath}
\usepackage{amssymb}

\usepackage{graphicx}

\newcommand{\vect}[1]{\ensuremath{\mathbf{#1}}}
\newcommand{\abs}[1]{\ensuremath{\left\lvert#1\right\rvert}}
\newcommand{\norm}[1]{\ensuremath{\lVert#1\rVert}}

\DeclareMathOperator*{\sinc}{sinc}

\begin{document}

\title{Performance of a GPU-based Direct Summation Algorithm for Computation of Small Angle Scattering Profile}

\author{
\IEEEauthorblockN{Konstantin Berlin\IEEEauthorrefmark{1}\IEEEauthorrefmark{2}\IEEEauthorrefmark{3}, Nail A. Gumerov\IEEEauthorrefmark{2}, Ramani Duraiswami\IEEEauthorrefmark{2},
David Fushman\IEEEauthorrefmark{1}\IEEEauthorrefmark{2}}\\
\IEEEauthorblockN{\IEEEauthorrefmark{1}Department of Chemistry and Biochemistry, \\Center for Biomolecular Structure and Organization,\\
 University of Maryland, College Park, MD 20742, USA}\\
\IEEEauthorblockN{\IEEEauthorrefmark{2}Institute for Advanced Computer Studies, \\
University of Maryland, College Park, MD 20742, USA}\\
\IEEEauthorblockN{\IEEEauthorrefmark{3}{Contact Email: kberlin@umd.edu}}}



\maketitle

\begin{abstract}

Small Angle Scattering (SAS) of X-rays or neutrons is an experimental technique that provides valuable structural information for biological macromolecules under physiological conditions and with no limitation on the molecular size. In order to refine molecular structure against experimental SAS data, \textit{ab initio} prediction of the scattering profile must be recomputed hundreds of thousands of times, which involves the computation of the $\sinc$ kernel over all pairs of atoms in a molecule. The quadratic computational complexity of predicting the SAS profile limits the size of the molecules and and has been a major impediment for integration of SAS data into structure refinement protocols.  In order to significantly speed up prediction of the SAS profile we present a general purpose graphical processing unit (GPU) algorithm, written in OpenCL, for the summation of the $\sinc$ kernel (Debye summation) over all pairs of atoms. This program is an order of magnitude faster than a parallel CPU algorithm, and faster than an FMM-like approximation method for certain input domains. We show that our algorithm is currently the fastest method for performing SAS computation for small and medium size molecules (around $50000$ atoms or less). This algorithm is critical for quick and accurate SAS profile computation of elongated structures, such as DNA, RNA, and sparsely spaced pseudo-atom molecules.
\end{abstract}


\section*{Introduction}

Accurate characterization of biomolecular structures in solution is required for understanding their biological function and for therapeutic applications. Small-angle scattering (SAS) of X-rays and neutrons indirectly measures distribution of interatomic distances in a molecule \cite{Kotch2003:QREV}. Unlike high-resolution techniques, such as X-ray crystallography and solution NMR, SAS allows the study of molecules and their interactions under native physiological conditions and with essentially no limitation on the size of the system under investigation. Though providing a powerful and unique set of experimental structural constraints, SAS is limited in resolution, as well as the ambiguity in deconvolution of interatomic distances from experimental data. It does not provide enough data to derive a high resolution molecular structure, but due to the ease of collecting SAS data at various conditions and the unique scale of atomic distance information, it is an extremely promising complement to the high-resolution techniques \cite{rambo2010bridging}. 

Solution SAS studies have become increasingly popular, with the applications covering a broad range, including structure refinement of biological macromolecules and their complexes \cite{Grishaev2008protein:JBM, Grishaev2008RNA:JBM, Grishaev2005:JACS, Pons2010:JMB}, and  analysis of conformational ensembles and flexibility in solution \cite{Bernado2010:BIO, Datta2009:JMB, Jehle2011:PNAS, Bernado2007:JACS}. Finally, SAS is starting to be used in high-throughput biological applications \cite{hura2009robust,grant2011saxs}.

In order to use SAS for structural refinement, the scattering profile of the SAS experiment needs to be predicted \textit{ab initio} from molecular structure, which requires computing all-pairs interactions of the atoms in the molecule (also referred to as an N-body problem). In addition, the profile must be recomputed hundreds of thousands of times in an iterative structure refinement algorithm, ensemble analysis \cite{Grishaev2008protein:JBM,grishaev2010improved}, or for thousands of different structures in a high-throughput method. For such applications serial computation of the scattering profile becomes prohibitive, even for smaller molecules.

Several approximation algorithms have been proposed to speed up this computation \cite{svergun1995crysol,bardhan2009softwaxs,grishaev2010improved,poitevin2011aquasaxs}. However, depending on the diameter of the molecule, the approximations can introduce significant errors (see \cite{gumerov2012hierarchical} for theoretical limitations of these algorithm). Recently a hierarchical harmonic expansion method based on the fast multipole method (FMM) has been shown to have superior asymptotical performance than all previously proposed approximation methods \cite{gumerov2012hierarchical}, while maintaining any prescribed accuracy. However, the algorithm is very complex, which makes it difficult to implement and parallelize.

Here we describe a parallelization of the direct computation of SAS profile onto readily affordable graphical processing cards (GPUs) that provides a dramatic improvement in the efficiency of the scattering profile prediction. While CUDA \cite{nvidia2007compute} GPU parallelization has been proposed for a similar problem of powder diffraction \cite{gelisio2010real}, we extend GPU parallelization to SAXS/WAXS/SANS using an open GPU programming standard, OpenCL \cite{khronos2008opencl}, that makes our software compatible with most modern desktop and laptop systems. We show that our GPU implementation is an order of magnitude faster than the parallelized CPU version, and faster than the parallelized CPU version of the hierarchical harmonic expansion.

\section{Background}

\subsection{Computational Complexity of SAS Scattering Profile}

The scattering profile is a function of the momentum transfer $q$,
\begin{equation}
I(q) =\left\langle \abs{\sum_{j=1}^{N}f_{j}(q) \exp \left( i\vect{q}\cdot \vect{r}_{j}\right)}^2 \right\rangle_{\Omega},
\label{eq:iq}
\end{equation}
where $N$ is the number of atoms in the molecule, $f_j$ is the atomic form factor of the $j$th atom, $\vect{r}_j$ is the position of the $j$th atom relative to some coordinate frame, $\left\langle \right\rangle_{\Omega}$ represents the averaging over all possible orientations of the molecule (this is mathematically equivalent to the averaging over all possible orientations of \vect{q}), with q=\norm{\vect{q}}. We can analytically integrate Eq. \eqref{eq:iq} over $\Omega$, which yields an all-pairs $\sinc$ kernel summation (Debye summation),
\begin{equation}
  I(q) = \sum_k^N f_k(q) \sum_j^N f_j(q) \frac{\sin(qr_{jk})}{qr_{jk}},
  \label{eq:debye}
\end{equation}
where $r_{jk}$ is the distance between the $j$th and $k$th atomic centers \cite{Debye15:APL,Kotch2003:QREV}. Unlike the case of powder diffraction \cite{gelisio2010real}, in the general SAS case, the form factors $f_j$ cannot be factored out of the equation, because the non-uniform density of the water layer atoms around the molecule creates an intractable number of possible $f_k f_j$ combinations.

The computational cost of directly evaluating Eq. \eqref{eq:debye} for each $q$ is $O(N^2)$. For larger molecular system (tens or hundreds of thousands of atoms), the quadratic computational cost is a prohibitive barrier for atomic level computation of Eq. \eqref{eq:debye}. This computational cost is further compounded in a structure refinement algorithms, when the Debye sum has to be repeatedly calculated as part of a structural refinement protocol, e.g., \cite{Lipfert07:ARBBS, Franke09:JAC,Grishaev2008RNA:JBM}, or if an ensemble of molecular conformations is considered. For example, on a modern desktop directly computing a scattering profile a single time could take minutes for a moderately sized $20000+$ atom molecule.

To address the quadratic complexity of Eq. \eqref{eq:debye} several papers have suggested approximately computing Eq. \eqref{eq:iq} by averaging over a constant number of orientations \cite{bardhan2009softwaxs,grishaev2010improved,poitevin2011aquasaxs}, or using harmonic expansion \cite{svergun1995crysol}. Assuming one accurately computes the averaged $I(q)$ value for some constant number of orientations or harmonic expansions, the algorithm should have a superior $O(N)$ complexity. However, it was demonstrated in \cite{gumerov2012hierarchical} that the number of orientations (or harmonics) needed for accurate orientational averaging is problem dependent, requiring no less than $O(q^2D^2)$ orientational samples, where $D$ is the the diameter of the molecule (the largest distance between two atoms of the molecule). Thus, the computation of the full profile has an overall complexity of $O(q^2D^2N)$. In addition, the constant inside the $O(q^2D^2N)$ is larger than in direct $O(N^2)$ computation. 

For a significant variety of structures (especially elongated structures like DNA or RNA), the factor $q^2D^2$ can be larger than $N$. For example, the complex [PDB:2R8S] has $12000$ atoms, and diameter of $150$ $\text {\AA}$. For $q=1.0$ $\text {\AA}^{-1}$, since $q^2D^2=150^2>N$, it clearly makes more sense to directly compute Eq. \eqref{eq:debye} for this complex, instead of using the approximations mentioned above. Similar analysis between direct and approximate methods can be done automatically for any type of input.

Recently we have demonstrated that the complexity of the approximate computation can be significantly improved by applying an FMM-like hierarchical algorithm based on spherical harmonics \cite{gumerov2012hierarchical}, giving an overall complexity of $O(N\log N+(qD)^3\log (qD))$ for a specific value of $q$. This method offers significant speedup in evaluation of Eq. \eqref{eq:iq} for most relevant ranges of $qD$ and $N$ as compared to previous approximation methods. 

However, due to the $(qD)^3$ dependence, as we demonstrate below, for medium size molecules and WAXS values of $q$ ($q>  0.25$), it is still faster to compute the $O(N^2)$ Eq. \eqref{eq:debye}, directly on the GPU. Our results clearly demonstrate a large computational benefit of a GPU based direct Debye summation algorithm for higher ranges of $qD$ and/or lower density pseudo-molecules \cite{Franke09:JAC,yang2009rapid}.

\subsection{GPU Computing}

GPUs with a substantial computing power are widely available today in most PCs, and can now provide over $1$ Tflop of computational power, such as the Tesla C2070.
While GPUs were originally designed for intense graphics computations, using GPUs as a less expensive alternative to the high-performance computing on a  CPU cluster is now a well established technique in scientific computing, and is commonly referred to as General Purpose computing on the GPU (GPGPU). Whereas a general purpose CPU devotes much of the on-chip real estate to flow-control and caching, and only a relatively small area to computational cores, GPUs have many cores that simultaneously execute the same small program on different data, referred to as Single-Instruction Multiple-Data (SIMD) architecture \cite{owens2008gpu}.

The GPGPU paradigm is well suited for structural biology because of the embarrassingly parallel nature of molecular force fields and the large number of particles in a typical molecular system. Such computations, running on the GPU, have been shown to give order of magnitude speedup over their parallel CPU companions (see \cite{stone2010gpu} for a review), but require the programmer to explicitly manage different
memory levels in order to achieve good performance.

The most popular method to program a GPU is with CUDA (interface developed by NVIDIA), which allows
programming of the GPUs in a C-like language. Detailed description of an efficient CUDA implementation for an N-body problem is provided by NVIDIA \cite{nyland2007fast} and more examples can be found, see \cite{stone2010gpu} for references. However, these implementations are available only for CUDA enabled NVIDIA cards, and cannot be used on GPUs from other vendors.

\begin{figure}[htb]
\centering
\includegraphics[width=3in]{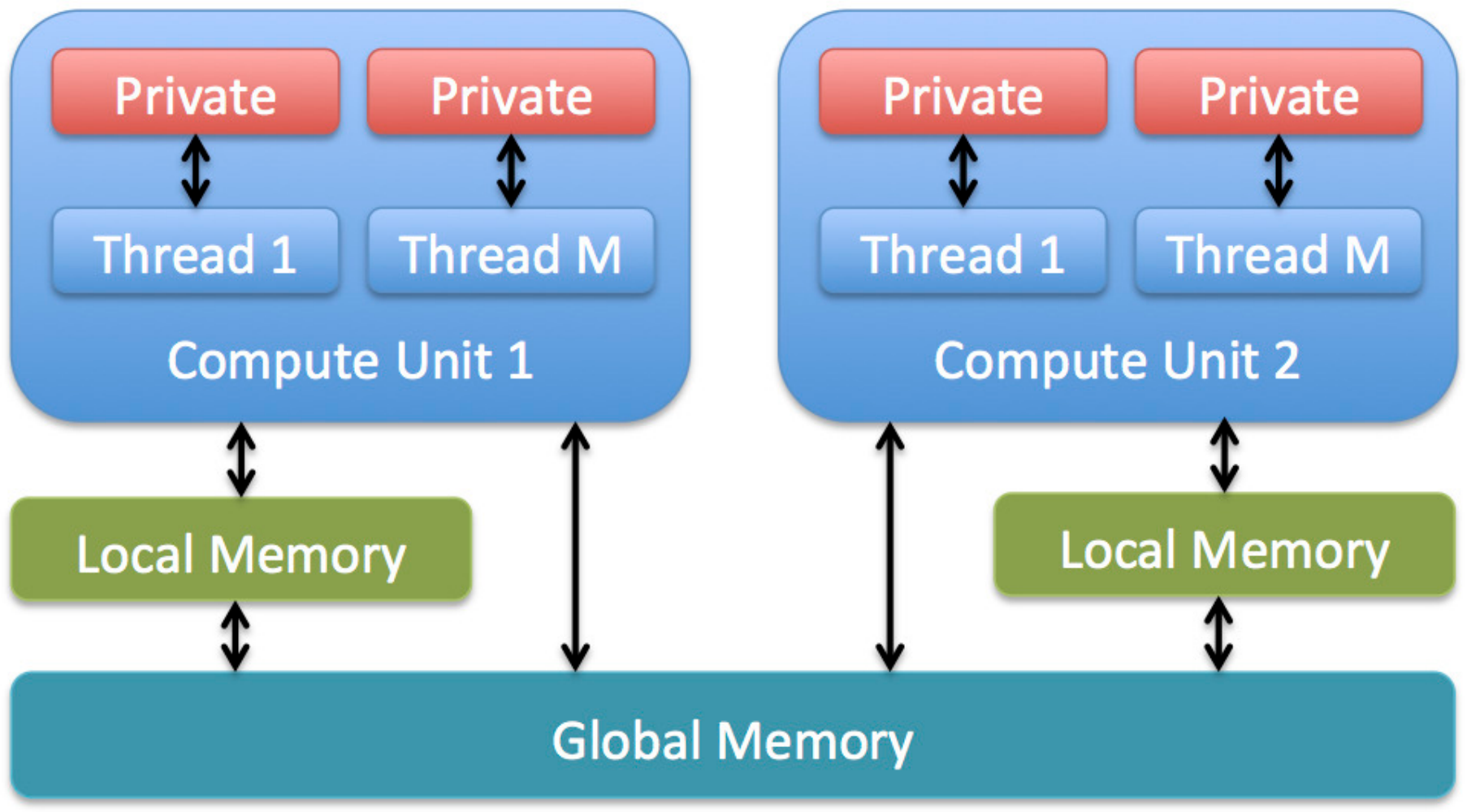}
\caption{Simplified OpenCL memory model, representing the three main levels of memory: global, local, and private. }
\label{fig:memory}
\end{figure}

To standardize and unify GPU programming across multiple platforms, the OpenCL standard was introduced in 2008 \cite{khronos2008opencl}. The diagram of the three basic OpenCL memory levels is given in Figure \ref{fig:memory}. GPU global memory is a large block of memory that is shared between all worker threads, however access to it is extremely slow relative to the local and private memory. Local memory is orders of magnitude faster than global memory, but is only shared between threads in an individual compute unit. Finally, each thread has a very small block of fast private memory. An OpenCL workgroup is a a subset of the compute unit, whose size is determined at compilation time.

In order to design an efficient GPU algorithm the majority of the computation in each thread should be done in private and local memory, with only an aggregate result periodically copied back to global memory. A corollary to this is that the amount of work done by each compute unit should be maximized relative to the amount of memory transfer between global and local/private memory. With this in mind, we now move on to describing our algorithm for the computation of the $\sinc$ kernel.

\section{Implementation}

We first observe than for each value of $q$, direct evaluation of Eq. $\eqref{eq:iq}$ is embarrassingly parallel, and can be significantly sped up by using parallel GPU computing. Given $O(N^2)$ work threads, the computation can be theoretically computed in $O(\log N)$ time and $O(N^2)$ work, with an $O(1)$ time parallel kernel evaluation, followed by an $O(\log N)$ time parallel reduction. In practice such an algorithm is not scalable because of the $O(N^2)$ memory requirement. Let $W<N^2$ be the number of workers available in the hardware of the GPU. We set out to design an $O(N^2/W+\log W)$ time algorithm with $O(N^2)$ work complexity, and $O(N+W)$ memory requirement. This algorithm scales linearly with $Q$, the number of $q$ values that need to be evaluated.

Since we expect users to run structure refinement computation on a heterogeneous set of systems, based on what is available to them at any given time (including personal laptops), we designed the algorithm for the general OpenCL 1.0 memory model. We focused on optimizing memory transfers between the RAM and global GPU memory, and between global GPU memory and local GPU memory, as well as avoiding quadratic scaling in memory usage.

Let $B$ be the number of workers in each compute unit, and assume that the global GPU memory is large enough to store all $N$ positions of atoms and at least one set of $QN$ associated form factors for a specific $q$ value. For local GPU memory we assume that it is large enough to store $B$ atom positions and $B$ form factors associated with the specific $q$ value. If $QN$ form factors are too large to fit into the main memory, we atomically split the GPU computation into multiple computations on a subset of $q$ values. Just like any implementation of the Debye summation, the algorithm consists of three loops: the loop over the $q$ values; the loop over all the atoms; and another loop over all the atoms.

The overwriting design of the algorithm is to minimize the number of memory transfers from global memory to local memory, and to avoid access of global memory outside of block transfers. In a normal CPU implementation the iteration over $q$ would be inside the double loop over the atoms, since this avoids recomputing the distance between two atoms at different values of $q$ (which requires an expensive square root operation). However, in our implementation we put the iteration over $q$ as the outer most loop due to the size limitation of local GPU memory, and only compute two $q$ values at a time inside the innermost iterations. Further increasing the number of the $q$ values does not improve the performance, since the storage of $f(q)$ values overwhelms the private and then the local memory, while also decreasing the maximum workgroup size.

\begin{figure}[htb]
\centering
\includegraphics[width=3in]{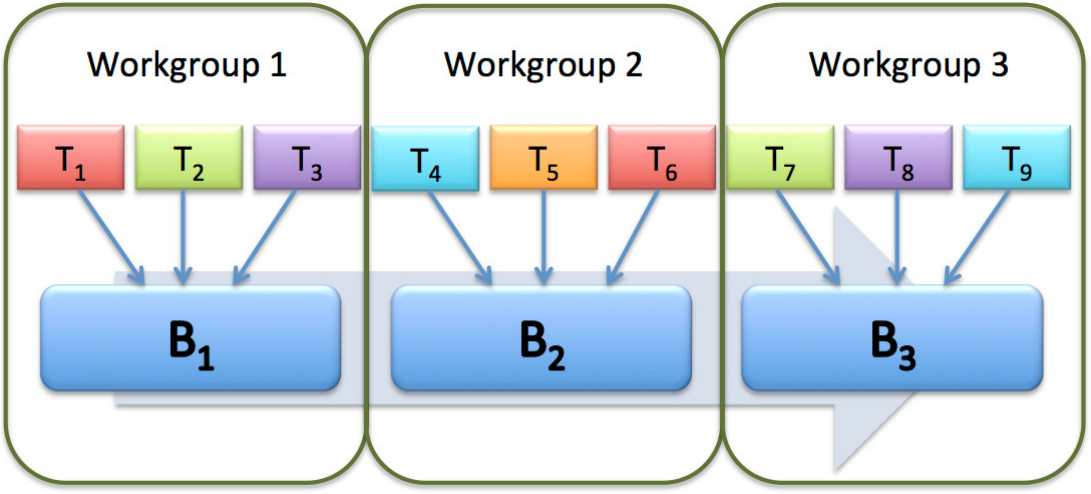}
\caption{ Diagram of our GPU algorithm. Each atom thread ($T$) computes the $\sinc$ kernel between itself and all the atoms in block $B$ (all the atoms are split into block, which are stored in local memory). After all compute units are done with their computation, the blocks are shifted by one, and the computation is repeated. }
\label{fig:process}
\end{figure}

The algorithm starts by spawning $N$ workers. For every second $q$ value, each worker $j$ copies its atomic position and the associated form factor from global memory to private memory and each local workgroup of size $B$ copies a block of $B$ sequential form factors and atom positions into the local memory. At this point each worker in the local workgroup proceeds to compute the kernel sum between its private atom and all the atoms stored in the local memory of the group. Once the computation is completed, the next block of $2B$ form factors (for the two $q$ values) and atom positions is copied into the local memory, and the computation is repeated. At the end of the all-pair summation for the two current $q$ values, a local parallel reduction is done to aggregate the private totals of each worker into one local memory value, and that value is copied into global memory. At the end of computation another reduction is done to aggregate all the workgroup totals, stored in global memory, into a final $I(q)$ array. The final reduction is a quick computation, and can be done in a CPU. In order to avoid unnecessary computation, only values for $j \leq k$ are computed. In the case when the number of atoms is smaller than the number of workers in a GPU, $N<W$, we create $pN \geq W$ worker, such that $(p-1)N>W$, and the internal loop over all atoms that was previously done by one worker, is now split between $p$ workers. The basic diagram of our algorithm is shown in Figure \ref{fig:process}.

Due to the limitation of some GPU cards, the OpenCL code was implemented using single instead of double precision. This does not affect the applicability of the computation, since the errors (even for a million atom system) are orders of magnitude below experimental errors and the inherent accuracy of the theoretical model. We expect the computation using doubles to be twice as slow, however it is unnecessary in this case.

\section{Results}

The GPU algorithm (``GPU'') was compared to three other methods: two methods are different implementations of the direct summation of Eq. \eqref{eq:debye}, IMP toolbox implementation in C++ (``IMP'') \cite{forster2008integration} and our CPU parallelized Java implementation (``Parallel''); and the other method is a parallelized CPU FMM-like hierarchical method \cite{gumerov2012hierarchical} (``FMM'', with the prescribed relative accuracy of $0.001$). We did not compare our method to other approximation methods since the FMM-like method has been demonstrated to have orders of magnitude faster performance than $O(q^2D^2N)$ approximation methods.

We performed benchmarks on a laptop system, as well as a significantly more expensive specialized computer. The laptop is a dual core 2.66 GHz Intel Core i7 Macbook Pro with NVIDIA GeForce GT 330M GPU, running Mac OS X Lion 10.7.2., C++ Apple LLVM compiler 4.2, CUDA 4.0, and Java 1.6. 
The second system is a single node of the Chimera GPU cluster, which has dual Intel Xeon quad-core 2.5 Ghz processors with Nvidia GTX 295 GPU, running Linux 2.6.18, Intel 11.1 compiler, and Java 1.6. 

All C++ and Fortran programs were compiled with the ``-O3'' flag. The harmonic expansion method was only run on the Chimera node due to compiler incompatibility on the Mac OS. Timing results were obtained for the critical region of the moment transfer values of $0<q\leq1.0$ $\text {\AA }$, as suggested in \cite{grishaev2010improved}.

\begin{figure}[htb]
\centering
\includegraphics[width=3in]{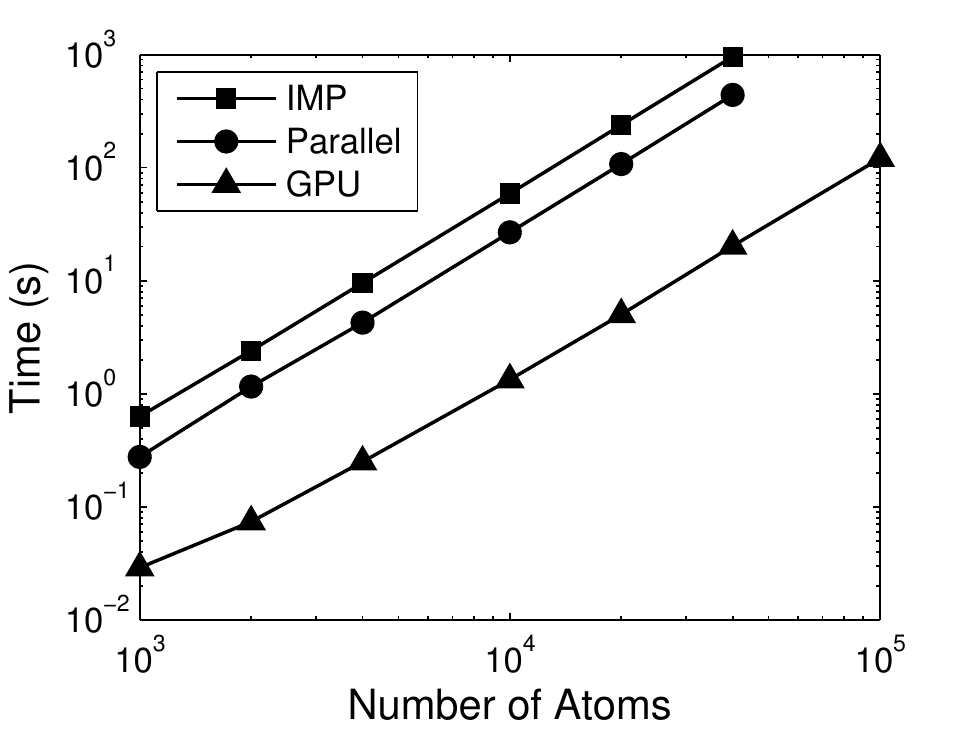}
\caption{ Timing results for the computation of a $50$ point SAS profile on differently sized, randomly generated proteins on the Macbook Pro system.}
\label{fig:macbookpro}
\end{figure}

The results for the Macbook Pro are shown in Figure \ref{fig:macbookpro}, and the detailed CPU and GPU timing results are also shown in Table \ref{tab:results}. The GPU version is over $47$ times faster than the standard single CPU computation in the IMP library, thus decreasing the computation time for a moderately large $40000$ atom protein from $957$ seconds to around $20$ seconds. For smaller proteins the speedup decreases due to the overhead of CPU-GPU memory transfers becoming a more prominent percent of the overall time. For a commonly sized $10000$ atom protein used in a SAS experiment, the GPU version was $44$ times faster with the time decreasing from $59$ seconds for IMP, to around $1.4$ seconds. 

\begin{table*}[!ht]
\centering
\caption{
Timing results of the described algorithms. }
\begin{tabular}{| l | c | c | c | c | c | c |} \hline
N & \multicolumn{3}{| c |}{Macbook Pro} &  \multicolumn{3}{| c |}{Chimera}\\ \hline
 & CPU (2 core) & GPU & Speedup & CPU (8 core) & GPU & Speedup\\ \hline
1000 & 0.28 & 0.03 & 9.5 & 0.10 & 0.02 & 5.9 \\ \hline
2000 & 1.16 & 0.07 & 15.6 & 0.35 & 0.03 & 13.0 \\ \hline
4000 & 4.25 & 0.25 & 17.0 & 1.35 & 0.08 & 17.3 \\ \hline
10000 & 26.8 & 1.34 & 20.1 & 8.28 & 0.34 & 24.3 \\ \hline
20000 & 108 & 5.02 & 21.5 & 33.8 & 1.14 & 29.6 \\ \hline
40000 & 439 & 20.4 & 21.6 & 133 & 3.94 & 33.7 \\ \hline
100000 & (2695) & 121 & 22.3 & 806 & 22.5 & 35.9 \\ \hline
\end{tabular}
\begin{center}{All timing results are in seconds. Estimated timing is shown in parentheses.}
\end{center}
\label{tab:results}
\end{table*}

Our parallel Java implementation was over $2$ times faster than the single threaded IMP implementations on our dual core system, with over $100$\% efficiency (this is most likely due to hyper-threading available on each core). The speedup of a GPU Debye summation over a similarly parallelized CPU version is over $22$ times on the Macbook Pro. Approximately $20$\% of the computation is taken up by memory transfers, $15$\% by the atomic distance computation, and the rest by evaluation of the $\sinc$ function and other unavoidable operations.

With such a dramatic speedup, the GPU algorithm makes it possible to perform previously infeasible SAS profile computations of moderate size proteins or even to compute the SAS profiles for a large ensemble of structures on a basic laptop.

\begin{figure}[htb]
\centering
\includegraphics[width=3in]{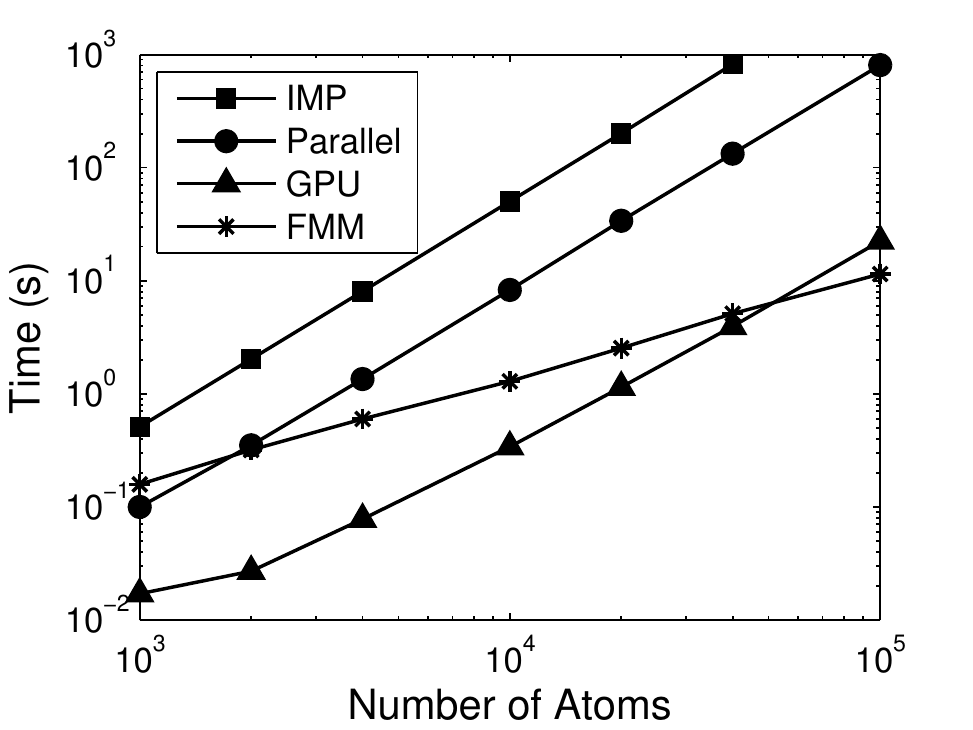}
\caption{Timing results for the computation of uniformly spaced $50$ point SAS profile with $0<q\leq1.0$ $\text {\AA }$ on different size randomly generated proteins on a single Chimera node.}
\label{fig:chimera}
\end{figure}

Similar results for a single Chimera node are shown in Figure \ref{fig:chimera}, and the detailed CPU and GPU timing results are also shown in Table \ref{tab:results}. The Chimera node is a significantly more expensive and powerful computer than the Macbook Pro, however for single threaded code, like IMP, the performance is only $1.15$ times faster than on the Macbook Pro. 

On the Chimera node, our parallel Java implementation is over $6$ times faster than IMP (below the perfect speedup of $8$, with about $94$\% parallelization efficiency), and our GPU implementation is around $36$ times faster than our parallel Java version for $100000$ atom protein, dropping to about $25$ times speedup for $10000$ atom protein. If we assume perfect parallelization of the C++ IMP code, the speedup drops to about $26$ for $100000$ atoms. This is similar to the speedup observed on the Macbook Pro. Complete timing results for the parallel CPU version and the GPU version on both systems are shown in Table \ref{tab:results}.

\begin{figure}[htb]
\centering
\includegraphics[width=3in]{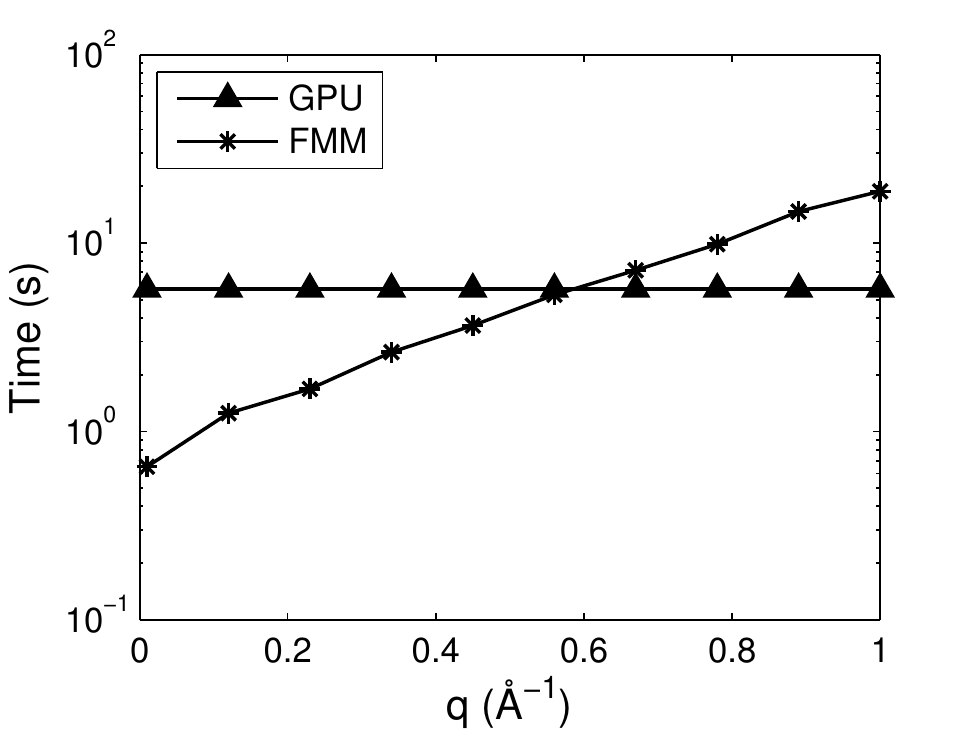}
\caption{Timing results at the individual $q$ values of a SAS profile of a randomly generated $50000$ atom protein, on a single Chimera node.}
\label{fig:qvalues}
\end{figure}

The parallelized CPU FMM-like algorithm becomes faster than the direct GPU computation at around $50000$ atoms. However, due to the dependence of FMM-like algorithm on $q$, the FMM-like algorithm is actually slower for higher $q$ values in the profile. In Figure \ref{fig:qvalues} we demonstrate the switch in the optimal computation algorithm for a $50000$ atom protein, from the FMM-like algorithm, to the GPU algorithm. The crossover between the FMM-like method and GPU is around $q=0.6$ $\text {\AA}^{-1}$, at which point our GPU implementation is still the fastest available method for computing a SAS profile. This result does not take into account the additional computational time of setting up the FMM, such as generating an octree and precomputing translation operators.

We demonstrate a practical application of our GPU method by comparing our {\it ab initio} prediction to the popular and extremely fast CRYSOL \cite{svergun1995crysol} software package using experimental data. The experimental data are for low $q$, where approximation methods, like CRYSOL, are most applicable, yet we still demonstrate a significant boost in computation speed.  We note that our algorithm does not have to be used as a standalone component, as is demonstrated here, but can be quickly integrated into a more advanced SAS pipeline that would yield improved fit to experimental data. Here we demonstrate the immediate computational advantage of our implementation, even for lower $q$ values, over the current state-of-the-art spherical harmonic approximation method.  

All hydrogens where explicitly treated and all {\it ab initio} prediction was done without refinement of any parameters against experimental data. CRYSOL was ran with default options, and no water layer was added to our method. Both, the profile for CRYSOL and GPU method was computed for $100$ uniformly distributed points. We provide the {\it ab initio} fits to experimental data solely to demonstrate that we are timing the necessary amount of experimentally relevant computation relative to CRYSOL. The final fits given in most publications include some amount of pre/post processing that is not relevant to our analysis.

\begin{figure}[htb]
\centering
\includegraphics[width=3in]{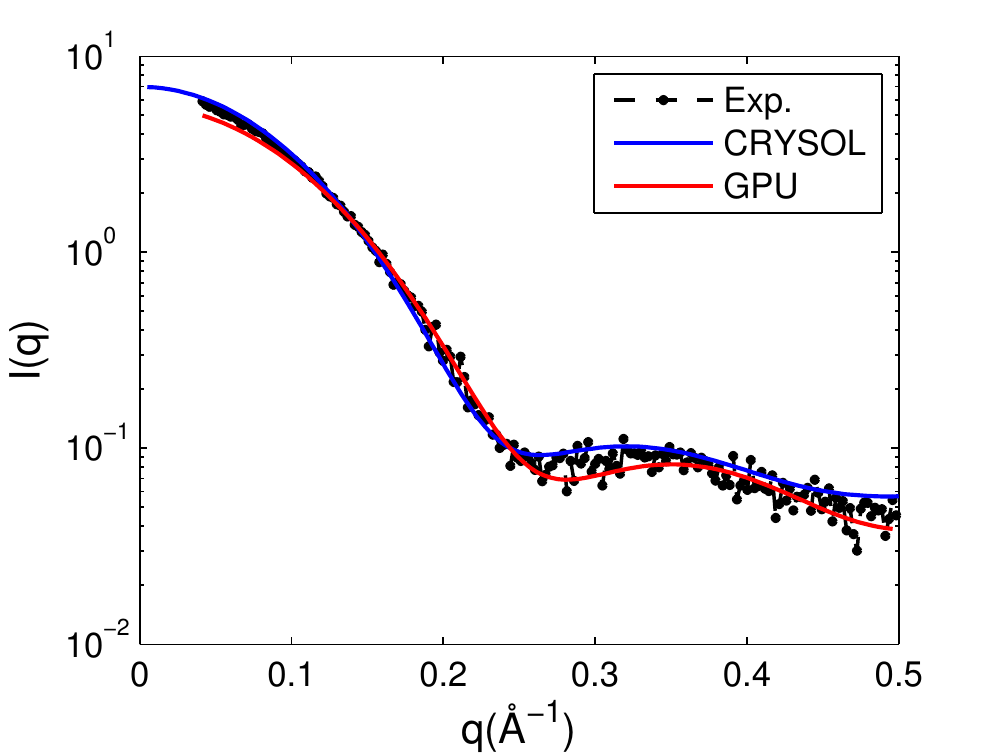}
\caption{Comparison of SAXS profile fit between CRYSOL and our GPU implementation for hen egg-white lysozyme [PDB:193L]. The predicted profiles were rescaled to overlap the experimental profile.}
\label{fig:lyzfit}
\end{figure}

The first benchmark was done on the lysozyme data provided in the CRYSOL package.  The computation was timed on the Macbook Pro machine, with the GPU computation taking $0.4$ seconds, and CRYSOL $1.0$ seconds. For visualization purposes, the profile fits are given in Figure \ref{fig:lyzfit}.

\begin{figure}[htb]
\centering
\includegraphics[width=3in]{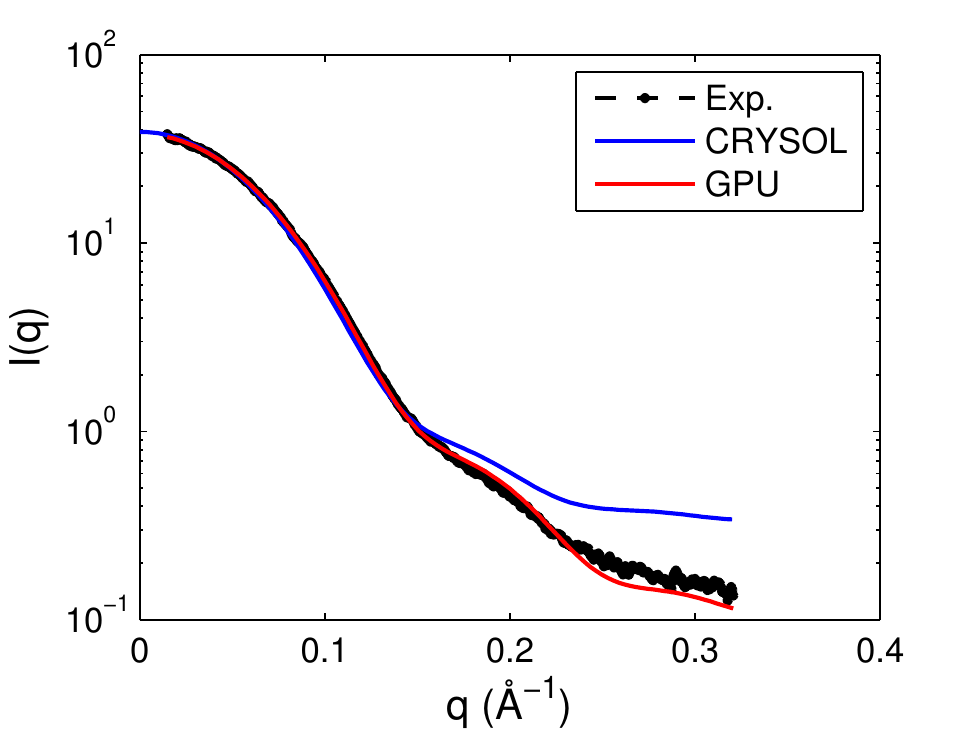}
\caption{Comparison of SAXS profile fit between CRYSOL and our GPU implementation for [PDB:3K3K]. The predicted profiles were rescaled to overlap the experimental profile.}
\label{fig:3K3Kfit}
\end{figure}

The second benchmark was ran on the dataset [BID:1PYR1P], downloaded from the BIOSIS database. The GPU method ran on a Macbook Pro in $1.2$ seconds, compared to $1.4$ seconds for CRYSOL. The fit is shown in Figure \ref{fig:3K3Kfit}.

The timing results demonstrate the advantage of our direct GPU algorithm, even at low $q$ values, over one of the fastest and most popular current implementations of approximation method.

\section{Discussion}

We have demonstrated that, for important input domains, our GPU implementation is currently the fastest possible algorithm for high $qD$ inputs, or small molecules, that can effectively run on a personal computer. This direct GPU implementation should be considered as a complementary algorithm to the CPU FMM-like implementation at lower $q$ values, where the algorithm choice should be made automatically, depending on the value of $qD$ and $N$. In a typical range of $0<q<1.0$ $\text {\AA}^{-1}$, we suggest that for low $q$ values an FMM-like algorithm should be used, while for higher $q$ values computation should be switched to the direct GPU algorithm. 

Since the CPU FMM-like and the GPU algorithms run on separate processors, both should run at the same time, on different $q$ ranges, in order to achieve maximum efficiency. The GPU algorithm performance should further improve in the future, as the size of the local memory increases and more $q$ values can fit inside the innermost loop. 

\section{Conclusion}

We provided the first comparison of a direct a GPU computation vs. FMM-like method for SAS problem, and demonstrates the immense utility of a parallel GPU direct summation method for a large set of potential input. At the same time, we showed the inherent computational tradeoff between approximation methods vs. direct all-to-all summation. Our implementation was written in OpenCL, allowing it to be run on most modern laptop and desktop GPUs, while still having over $25$ times faster runtime than a similar parallel CPU implementation.

We also demonstrated an application of our new method on two experimental datasets, and showed that our GPU software can compute the scattering profile faster than CRYSOL, with a guaranteed machine level computational precision of the underlying mathematical model. The performance gains of this algorithm can be quickly realized by any of the current {\it ab initio} SAS prediction packages.

The above software is part of a new high performance software package, ARMOR, which includes additional NMR restraints \cite{patidock,elmdock}.

\section*{Acknowledgements}

We thank the UMIACS New Research Frontiers Award for supporting this research.
 
\bibliographystyle{elsarticle-num}  
\bibliography{gpusaxs}    

\begin{thebibliography}{10}
\expandafter\ifx\csname url\endcsname\relax
  \def\url#1{\texttt{#1}}\fi
\expandafter\ifx\csname urlprefix\endcsname\relax\def\urlprefix{URL }\fi
\expandafter\ifx\csname href\endcsname\relax
  \def\href#1#2{#2} \def\path#1{#1}\fi

\bibitem{Kotch2003:QREV}
M.~Koch, P.~Vachette, D.~Svergun, Small-angle scattering: a view on the
  properties, structures and structural changes of biological macromolecules in
  solution, Quarterly Reviews of Biophysics 36~(02) (2003) 147--227.

\bibitem{rambo2010bridging}
R.~Rambo, J.~Tainer, Bridging the solution divide: comprehensive structural
  analyses of dynamic {RNA}, {DNA}, and protein assemblies by small-angle
  {X}-ray scattering, Current Opinion in Structural Biology 20~(1) (2010)
  128--137.

\bibitem{Grishaev2008protein:JBM}
A.~Grishaev, V.~Tugarinov, L.~Kay, J.~Trewhella, A.~Bax, Refined solution
  structure of the {82-kDa} enzyme malate synthase {G} from joint {NMR} and
  synchrotron {SAXS} restraints, Journal of Biomolecular NMR 40~(2) (2008)
  95--106.

\bibitem{Grishaev2008RNA:JBM}
A.~Grishaev, J.~Ying, M.~Canny, A.~Pardi, A.~Bax, Solution structure of {tRNA}
  {Val} from refinement of homology model against residual dipolar coupling and
  {SAXS} data, Journal of Biomolecular NMR 42~(2) (2008) 99--109.

\bibitem{Grishaev2005:JACS}
A.~Grishaev, J.~Wu, J.~Trewhella, A.~Bax, Refinement of multidomain protein
  structures by combination of solution small-angle {X}-ray scattering and
  {NMR} data, Journal of the American Chemical Society 127~(47) (2005)
  16621--16628.

\bibitem{Pons2010:JMB}
C.~Pons, M.~D'Abramo, D.~Svergun, M.~Orozco, P.~Bernad{\'o},
  J.~Fern{\'a}ndez-Recio, Structural characterization of protein-protein
  complexes by integrating computational docking with small-angle scattering
  data, Journal of Molecular Biology 403~(2) (2010) 217--230.

\bibitem{Bernado2010:BIO}
P.~Bernad{\'o}, K.~Modig, P.~Grela, D.~Svergun, M.~Tchorzewski, M.~Pons,
  M.~Akke, Structure and dynamics of ribosomal protein {L12}: an ensemble model
  based on {SAXS} and {NMR} relaxation, Biophysical Journal 98~(10) (2010)
  2374--2382.

\bibitem{Datta2009:JMB}
A.~Datta, G.~Hura, C.~Wolberger, The structure and conformation of
  {Lys63}-linked tetraubiquitin, Journal of Molecular Biology 392~(5) (2009)
  1117--1124.

\bibitem{Jehle2011:PNAS}
S.~Jehle, B.~S. Vollmar, B.~Bardiaux, K.~K. Dove, P.~Rajagopal, T.~Gonen,
  H.~Oschkinat, R.~E. Klevit, N-terminal domain of {$\alpha$B}-crystallin
  provides a conformational switch for multimerization and structural
  heterogeneity, Proceedings of the National Academy of Sciences.

\bibitem{Bernado2007:JACS}
P.~Bernad{\'o}, E.~Mylonas, M.~Petoukhov, M.~Blackledge, D.~Svergun, Structural
  characterization of flexible proteins using small-angle {X}-ray scattering,
  Journal of the American Chemical Society 129~(17) (2007) 5656--5664.

\bibitem{hura2009robust}
G.~Hura, A.~Menon, M.~Hammel, R.~Rambo, F.~Poole~II, S.~Tsutakawa,
  F.~Jenney~Jr, S.~Classen, K.~Frankel, R.~Hopkins, et~al., Robust,
  high-throughput solution structural analyses by small angle {X}-ray
  scattering ({SAXS}), Nature Methods 6~(8) (2009) 606--612.

\bibitem{grant2011saxs}
T.~D. Grant, J.~R. Luft, J.~R. Wolfley, H.~Tsuruta, A.~Martel, G.~T.
  Montelione, E.~H. Snell, Small angle {X}-ray scattering as a complementary
  tool for high-throughput structural studies, Biopolymers 95~(8).

\bibitem{grishaev2010improved}
A.~Grishaev, L.~Guo, T.~Irving, A.~Bax, Improved fitting of solution {X}-ray
  scattering data to macromolecular structures and structural ensembles by
  explicit water modeling, Journal of the American Chemical Society 132~(44)
  (2010) 15484--15486.

\bibitem{svergun1995crysol}
D.~Svergun, C.~Barberato, M.~Koch, {CRYSOL}-a program to evaluate {X}-ray
  solution scattering of biological macromolecules from atomic coordinates,
  Journal of Applied Crystallography 28~(6) (1995) 768--773.

\bibitem{bardhan2009softwaxs}
J.~Bardhan, S.~Park, L.~Makowski, {SoftWAXS}: a computational tool for modeling
  wide-angle {X}-ray solution scattering from biomolecules, Journal of Applied
  Crystallography 42~(5) (2009) 932--943.

\bibitem{poitevin2011aquasaxs}
F.~Poitevin, H.~Orland, S.~Doniach, P.~Koehl, M.~Delarue, {AquaSAXS}: a web
  server for computation and fitting of {SAXS} profiles with non-uniformally
  hydrated atomic models, Nucleic Acids Research 39~(suppl 2) (2011)
  W184--W189.

\bibitem{gumerov2012hierarchical}
N.~A. Gumerov, K.~Berlin, D.~Fushman, R.~Duraiswami, A hierarchical algorithm
  for fast debye summation with applications to small angle scattering, Journal
  of Computational Chemistry 33~(25) (2012) 1981--1996.

\bibitem{nvidia2007compute}
Nvidia, Compute unified device architecture programming guide (2007).

\bibitem{gelisio2010real}
L.~Gelisio, C.~Azanza~Ricardo, M.~Leoni, P.~Scardi, Real-space calculation of
  powder diffraction patterns on graphics processing units, Journal of Applied
  Crystallography 43~(3) (2010) 647--653.

\bibitem{khronos2008opencl}
A.~Munshi, et~al., The {OpenCL} Specification (2008).

\bibitem{Debye15:APL}
P.~Debye, Zerstreuung von r{\"o}ntgenstrahlen, Annalen der Physik 351~(6)
  (1915) 809--823.

\bibitem{Lipfert07:ARBBS}
J.~Lipfert, S.~Doniach, Small-angle {X}-ray scattering from {RNA}, proteins,
  and protein complexes, Annual Reviews of Biophysics and Bimolecular Structure
  36 (2007) 307--327.

\bibitem{Franke09:JAC}
D.~Franke, D.~Svergun, {DAMMIF}, a program for rapid ab-initio shape
  determination in small-angle scattering, Journal of Applied Crystallography
  42~(2) (2009) 342--346.

\bibitem{yang2009rapid}
S.~Yang, S.~Park, L.~Makowski, B.~Roux, A rapid coarse residue-based
  computational method for {X}-ray solution scattering characterization of
  protein folds and multiple conformational states of large protein complexes,
  Biophysical Journal 96~(11) (2009) 4449--4463.

\bibitem{owens2008gpu}
J.~Owens, M.~Houston, D.~Luebke, S.~Green, J.~Stone, J.~Phillips, {GPU}
  computing, Proceedings of the IEEE 96~(5) (2008) 879 --899.

\bibitem{stone2010gpu}
J.~E. Stone, D.~J. Hardy, I.~S. Ufimtsev, K.~Schulten, {GPU}-accelerated
  molecular modeling coming of age, Journal of Molecular Graphics and Modelling
  29~(2) (2010) 116--25.

\bibitem{nyland2007fast}
L.~Nyland, J.~Prins, Fast n-body simulation with {CUDA}, Simulation 3 (2007)
  677--696.

\bibitem{forster2008integration}
F.~F{\"o}rster, B.~Webb, K.~A. Krukenberg, H.~Tsuruta, D.~A. Agard, A.~Sali,
  Integration of small-angle {X}-ray scattering data into structural modeling
  of proteins and their assemblies, Journal of Molecular Biology 382~(4) (2008)
  1089--1106.

\bibitem{patidock}
K.~Berlin, D.~P. O'Leary, D.~Fushman, Structural assembly of molecular
  complexes based on residual dipolar couplings, Journal of the American
  Chemical Society 132~(26) (2010) 8961--8972.

\bibitem{elmdock}
K.~Berlin, D.~P. O'Leary, D.~Fushman, Fast approximations of the rotational
  diffusion tensor and their application to structural assembly of molecular
  complexes, Proteins: Structure, Function, and Bioinformatics 79~(7) (2011)
  2268--2281.

\end{thebibliography}


\end{document}